\documentclass[]{article}
\usepackage{emulateapj}
\usepackage{graphicx}

\advance \voffset by -.5cm\relax

\def\msun{{\rm M}_{\odot}}

\def\mpy{{\rm M}_{\odot} {\rm ~yr}^{-1}}

\begin{document}

\title{On The Origin Of The Highest Redshift Gamma-Ray Bursts}

 \author{Krzysztof Belczynski\altaffilmark{1,2}, 
         Daniel E. Holz\altaffilmark{1},
         Chris L. Fryer\altaffilmark{1}, 
         Edo Berger\altaffilmark{3},
         Dieter H. Hartmann\altaffilmark{4}, 
         Brian O'Shea\altaffilmark{5}
 }  

 \affil{
     $^{1}$ Los Alamos National Laboratory, Los Alamos, NM 87545, USA
            ({\em Oppenheimer Fellow}) \\ 
     $^{2}$ Astronomical Observatory, University of Warsaw, Al. Ujazdowskie 4, 
            00-478 Warsaw, Poland\\
     $^{3}$ Harvard-Smithsonian Center for Astrophysics, 60 Garden St., Cambridge, MA 02138, USA \\
     $^{4}$ Dept. of Physics and Astronomy, Clemson University, Clemson, SC 29634-0978, USA\\
     $^{5}$ Dept. of Physics and Astronomy, Michigan State University, East
            Lansing MI 48824, USA\\
     kbelczyn@nmsu.edu,daniel@restmass.com,clfreyer@lanl.gov,\\
     eberger@astro.princeton.edu,hdieter@CLEMSON.EDU,oshea@msu.edu
 }

\begin{abstract} 
GRB 080913 and GRB 090423 are the
most distant gamma-ray bursts (GRBs) known to-date, with spectroscopically
determined redshifts of $z=6.7$ and $z=8.1$, respectively. The detection of
bursts at this early epoch of the Universe significantly constrains the nature
of GRBs and their progenitors. We perform
population synthesis studies of the formation and evolution of early stars, and
calculate the resulting formation rates of short and long-duration GRBs at high
redshift.  The peak of the GRB rate from Population II stars occurs at $z \sim
7$ for a model with efficient/fast mixing of metals, while it is found at $z
\sim 3$ for an inefficient/slow metallicity evolution model. We show that in the
redshift range $6 \lesssim z \lesssim 10$ essentially all GRBs originate from
Population II stars, regardless of metallicity evolution model.
These stars (having small, but non-zero metallicity) are the most likely progenitors 
for {\em both}\/ long GRBs (collapsars) and short GRBs (NS-NS or BH-NS mergers) at 
this epoch. Although the predicted intrinsic rates of long and short
GRBs are similar at these high redshifts, observational selection effects lead
to higher (factor of $\sim 10$) observed rates for long GRBs.
We conclude that the two recently observed high-$z$ GRB events are most likely long 
GRBs originating from Population II collapsars.
\end{abstract}


\section{Introduction}

The {\em Swift} satellite has recently discovered
two high-redshift GRBs: GRB 080913 at redshift $z=6.7$ (Schady et
al. 2008), and GRB 090423 at redshift $z=8.1\mbox{--}8.3$ (Tanvir et al. 2009,
Salvaterra et al. 2009).  The gamma-ray properties of these bursts
straddle the traditional dividing lines between short and long bursts,
making a solid classification of each burst difficult.  For example,
the observed burst duration of GRB 080913, $\sim 8\mbox{ s}$,
classifies it as a long GRB (Stamatikos et al. 2008).  However, in the
rest frame of the burst the duration is $\sim 1\mbox{ s}$,
suggesting a short GRB (Perez-Ramirez et al. 2008). It has been argued
that, despite its relatively short duration (in the comoving frame),
GRB 080913 nonetheless should be classified as a long GRB due to the
specifics of the category definition (which places the dividing line
at $2\mbox{ s}$ in the observer frame, for a GRB sample with mostly
unknown redshifts), and because other properties of this burst (e.g.,
the lag-luminosity and Amati (2006) relations) are consistent with
those of long GRBs (Greiner et al. 2008).  Similarly, GRB 090423 
had a burst duration of $\sim 9\mbox{ s}$, and a rest-frame duration
of $\sim 1\mbox{ s}$. Although many
of the gamma-ray diagnostics of this burst (T$_{90}$, E$_{\rm peak}$, hardness ratio)
would classify it as ``short'' had the burst occurred at low
redshift, it has been argued that GRB 090423, as in the case of GRB 080913, is a
``long'' burst (Salvaterra et al. 2009; Zhang et al. 2009).

We employ two simplifications of nomenclature. First, we retain the
use of the ``short'' and ``long'' classifications for GRBs, despite
growing evidence that the classification should be determined using a
broader set of criteria beyond merely burst duration (Donaghy et
al. 2006; Zhang et al. 2007; Bloom et al.~2008). Second, we tie these
two classifications to specific progenitors using the simple theory
correspondences outlined in Popham et al.~(1999).  Long GRBs are
believed to be associated with the deaths of massive stars, and the
associated formation of black holes (e.g., Woosley 1993; Galama et
al. 1998; Hjorth et al. 2003; Stanek et al. 2003; Woosley \& Bloom
2006). Short GRBs, on the other hand, are thought to originate from
the merger of compact objects, such as double neutron star (NS-NS) or
black hole neutron star (BH-NS) binaries (e.g., Paczynski 1986;
Eichler et al. 1989; Nakar 2007 and references therein). An alternative,
unified model for GRBs has been proposed to simultaneously explain
both long and short GRBs in terms of the same underlying physical
mechanism, involving a black hole as central engine (e.g., Ruffini
et al. 2006; Dar \& Rujula 2004). In addition, there exist models for
both short and long GRBs that do not involve black holes (e.g., Usov
1992; King, Olsson, \& Davies 2007; Cheng \& Dai 1996).  A preferable
classification of bursts might focus on the progenitor 
classification of each bursts, which depends upon supernova association, 
host galaxy type, metallicity, and offset of the GRB with respect to the 
host galaxy (see Fryer et al. 2007 for a review).  In this paper we study 
the progenitors directly, and identify collapsars with long GRBs, and compact
object mergers with short GRBs.

Given their cosmological nature, and their bright afterglows, GRBs are unique 
tools to probe cosmological parameters and the structure and composition of matter
along the GRB sightlines (e.g., Lamb \& Reichart 2000; Lloyd-Ronning et al. 2002; 
Gou et al. 2004; Prochaska et al., 
2007; Lamb 2007; Hartmann 2008; Savaglio et al. 2008; Piro et al. 2008; Hartmann
et al. 2009).
This argument gains force with the discovery of GRBs at very high redshifts; for
example GRB 050904 at $z=6.29$ (Cusumano et al. 2006; Kawai et al. 2006), and now
GRB 080913 and GRB 090423 beyond the furthest quasars (e.g., CFHQS J2329-0301 at $z=6.4$; 
Willot et al. 2007) and the furthest spectroscopically confirmed galaxies
(galaxy IOK-1; J132359.8+272456, at $z=6.96$; Iye et al. 2006). These
high redshift objects provide a unique opportunity to study the Universe during
the extended, inhomogeneous epoch of reionization, tracing cosmic star formation
back to the first generation of stars (Population III). 
GRB follow-up observations have the potential to reveal the fraction of neutral
hydrogen and metal abundances as a function of redshift, possibly beyond $z \sim 10$,
with future instrumentation on large aperture telescopes on the ground and in
space. GRBs constrain star formation models in the pristine Universe, and serve
as unique light sources to pinpoint star formation activity even in the smallest 
proto-galaxies which would otherwise remain undetected. Rapid absorption
spectroscopy of their afterglows offers a truly exceptional opportunity
for studies of cosmic chemical evolution in an otherwise inaccessible part of the 
cosmic baryon field. However, in order to utilize these distant GRBs as probes,
prompt localization and redshift determination is crucial. Currently, the {\em Swift}\/ 
satellite performs rapid response observations in the X-ray and optical-UV bands. 
Future missions, such as the Energetic X-ray Imaging Survey Telescope (EXIST; Grindlay 
2006), will provide combined wide-field X-ray and near-infrared imaging spectroscopy 
to study distant GRBs and associated black hole formation in the early Universe.  

In this study we attempt to characterize the progenitors of GRB 080913 and GRB 090423,
and other potential GRBs at these high redshifts.  As we shall see, this epoch
is entirely dominated by Population II stars and we focus on GRBs
produced only from these stars.  We analyze the formation of
stars at high redshift in \S\,2.  In \S\,3 we use a self-consistent
model to follow in detail the evolution of the most likely GRB
progenitors: massive stars and NS-NS/BH-NS mergers. In \S\,4 we adopt
a cosmological model to estimate the GRB intrinsic (\S\,5) and observed
(\S\,6) event rates at high redshift, explicitly identifying the likely 
progenitors of GRB 080913 and 090423. We conclude in \S\,7.

\section{Stellar Populations: Model/Results}

{\em Star Formation History.}
To derive rates for various stellar events (supernovae, mergers, etc.) one must
start from a star formation rate history, and add assumptions about the cosmic 
evolution of basic properties such as the Initial Mass Function (IMF) and the 
fraction of stars in binaries. We use the analytic cosmic star formation 
rate (SFR) prescription provided by Strolger et al. (2004):
\begin{equation}
\mbox{sfr}(t) = 10^9 a (t^b e^{-t/c} + d e^{d(t-t_0)/c})\  
\,\msun\ {\rm yr}^{-1}\ {\rm Gpc}^{-3}
,
\label{eq1}
\end{equation}
where $t$ is the age of Universe (in Gyr) as measured in the global comoving (rest) frame,
$t_0$ is the present age of the Universe ($t_0=13.47$ Gyr; see \S\,4), and we adopt
parameters corresponding to the extinction-corrected model: $a=0.182$,
$b=1.26$, $c=1.865$, $d=0.071$. The SFR rate density described above is in comoving units 
(space as well as time); for a non-evolving population, this rate is constant.

{\em Galaxy Mass Distribution.}
At redshift $z<4$ we describe the distribution of galaxy masses using
a Schechter-type probability density function, calibrated to observations from 
Fontana et al. (2006):
\begin{equation}
\Phi(M_{\rm gal},z)= \Phi^{*}(z) \ln(10) a^{1+\alpha(z)} e^{-a}
\label{eq2}
\end{equation}
with
$\Phi^{*}=0.0035 (1+z)^{-2.20}$, $a=10^{\log(M_{\rm gal})-M_{z}}$, 
$M_{z}=11.16+0.17z-0.07z^2$, and $\alpha(z)=-1.18-0.082z$.
A galaxy mass, $M_{\rm gal}$, in units of $\msun$, is drawn from this
distribution (Eq.~\ref{eq2}) in the range $7<\log(M_{\rm gal})<12$.
For galaxies at redshifts beyond $z=4$ we use the above distribution evaluated 
at $z=4$, i.e., we assume no evolution in the galaxy mass distribution 
at earlier times. This assumption reflects the lack of information on galaxy mass
distributions at high redshift. Although this may be a decent approximation
at intermediate redshifts ($4\lesssim z \lesssim 10$), it is almost certainly
incorrect at very high redshift, where only low mass galaxies 
(and Population III stars) are forming. 

{\em Galaxy Metallicity.}
We assume that the average oxygen to hydrogen number 
ratio ($F_{\mbox{{\tiny OH}}}$) of a typical galaxy depends on its mass as:
\begin{equation}
\begin{array}{ll} 
\log(F_{\mbox{{\tiny OH}}})= & \log(10^{12} \mbox{O}/\mbox{H}) = \\ 
           & sz + 1.847 \log(M_{\rm gal})-0.08026 (\log(M_{\rm gal}))^2 \\
\end{array}
\label{eq3}
\end{equation} 
with a redshift independent normalization $sz=-1.492$ (Tremonti et al.~2004). 
It has been suggested that the functional form of this mass-metallicity relation
is redshift independent (Erb et al. 2006; Young \& Fryer 2007), with only the 
normalization factor, $sz$, varying with redshift. We characterize the redshift 
dependence of galaxy metallicities using the average metallicity relation from 
Pei, Fall, \& Hauser (1999):
\begin{equation}
Z \propto \left\{ \begin{array}{ll} 
10^{-a_2z}  & z<3.2 \\
10^{-b_1-b_2z}  & 3.2\leq z < 5 \\
10^{-c_1-c_2z}  & z \geq 5 \\
\end{array}
\right.
,
\label{eq4}
\end{equation} 
which implies evolution of the normalizing factor, $sz$, with redshift given by
\begin{equation}
sz \propto \left\{ \begin{array}{l} 
-a_2z-1.492  \quad\quad\quad\quad\quad\quad\quad\quad\quad\quad\quad\quad\quad z<3.2 \\
-b_2z-3.2(a_2-b_2)-1.492  \quad\quad\quad\quad\quad 3.2\leq z < 5 \\
-c_2z-5(b2-c2)-3.2(a2-b2)-1.492  \quad\quad z \geq 5 \\
\end{array}
\right.
\label{eq5}
\end{equation} 
With this approach we assume that the oxygen abundance (used in $F_{\mbox{{\tiny OH}}}$) 
correlates linearly with the average abundance of elements heavier
than Helium (which is what is provided in the metallicity measure, $Z$). 
The normalization given by Pei et al. (1999) is 
$a_2=0.5,\ b_1=0.8,\ b_2=0.25,\ c_1=0.2,\ c_2=0.4$. 
We also derive the coefficients from the metallicity dependence proposed by 
Young \& Fryer (2007), based on ultraviolet (GALEX, Sloan Digital
Sky Survey), infrared (Spitzer), and neutrino (Super-Kamiokande) observations 
(Hopkins \& Beacom 2006): 
$a_2=0.12,\ b_1=-0.704,\ b_2=0.34,\ c_1=0.0,\ c_2=0.1992$. 
The Pei et al. (1999) normalization results in a relatively slow metallicity
evolution as compared with the model of Young \& Fryer (2007), which allows for 
a rapid increase of the average metallicity of stars with time. This is shown
in Figure~\ref{sfr1} (top panel), which provides the metallicity distribution of
stars. It is apparent that for the {\em fast}\/ model the metallicity of
stars is on average an order of magnitude higher than for the {\em slow}\/
model at the same high ($z=7$) redshift.

Using Equations~\ref{eq3} and~\ref{eq5}, we estimate the average galaxy
oxygen to hydrogen ratio, given the mass and redshift of a galaxy. We express
our estimators relative to solar:
\begin{equation}
F_{\mbox{{\tiny OH},gal}}=F_{\mbox{{\tiny OH}}}/F_{\mbox{{\tiny OH},}\odot},
\label{eq6}
\end{equation}
where $F_{\mbox{{\tiny OH},}\odot}=4.9 \times 10^8$ is the solar value 
(Tremonti et al. 2004). 

{\em Galaxy Stellar Population.}
We use the above galaxy metallicity estimator to formally delineate different 
stellar populations:
\begin{equation}
 \begin{array}{ll}
F_{\mbox{{\tiny OH},gal}}<10^{-4}  & \mbox{Population III} \\
10^{-4} \leq F_{\mbox{{\tiny OH},gal}} \leq 10^{-1} & \mbox{Population II} \\
F_{\mbox{{\tiny OH},gal}}>10^{-1}  & \mbox{Population I} \\
\end{array}
.
\label{eq7}
\end{equation}
Our choice for the lower metallicity bound ($10^{-4}$) on Population II  marks the 
point where metals are abundant enough to provide sufficient cooling in the collapse 
of gas clouds, and thus star formation significantly deviates from the Population III 
stage (e.g., Mackey, Bromm, \& Hernquist 2003; Smith et al. 2008). It also marks the 
point where winds of massive stars are sufficiently strong to prevent the formation of 
pair-instability supernovae (Heger et al. 2003).  
The choice for the Population II upper bound is dictated by the observation of 
Population I and II stars in the Milky Way (e.g., Binney \& Merrifield 1998; Beers \&
Christlieb 2005). Population I stars (e.g., disk and bulge stars) have metallicities that are
approximately solar, with variations of a factor of five. Population II stars (e.g., halo 
populations, such as stars in globular clusters) have metallicities at or below 
$10^{-1}\mbox{--}10^{-2}$ solar. 

Each galaxy is assumed to host just one stellar population, determined by its
metallicity. Since we draw a large number of galaxies at any given redshift 
(see Eq.~\ref{eq2}) via Monte Carlo simulations, the use of an average 
metallicity measure is appropriate. 

In Figure~\ref{sfr1} (bottom panel) we show our assumed star formation history as a 
function of redshift, with the contribution from Population II stars highlighted.
Population II stars form over a wide range of redshifts, starting at $z \sim 22$, 
and peaking at $z \sim 7$ for the fast metallicity evolution model. 
For the slow metallicity evolution, the interstellar medium is enhanced with
metals much later, and Population II stars begin forming only at redshifts 
$z \sim 11$, peaking at $z \sim 3$. 
Although the metallicity evolution is poorly constrained, the two adopted
models are most likely extremes, with the true evolution somewhere
in between. For example, the transition from Population III to Population II
stars is found to occur at $z \sim 11$ and $z \sim 22$ for our slow and fast 
metallicity evolution models, respectively. More sophisticated models place this 
limit at redshifts $z \sim 15\mbox{--}20$ (e.g., Mackey et al. 2003).  
Recent results show that Population III stars can form as recently as
$z \sim 5$ (Schneider et al. 2006; Tornatore, Ferrara, Schneider 2007), but 
with rates significantly below that of Population II stars.

Population III stars were forming prior to the measured redshifts of GRB 080913
and 090423. Individual Population III stars may give rise to long GRBs in collapsar
models with enhanced rotation (e.g., Yoon, Langer, \& Norman 2006). However, as
the lifetime of massive Population III stars is very short ($\sim \mbox{3--6}$ Myr; e.g.,
Schaerer 2002), these GRBs would be observed at much higher redshifts ($z
\gtrsim 10$), tracing the Population III star formation rate with little delay.  
If binary Population III stars are formed, they could evolve into NS-NS/BH-NS systems 
that in turn, after an appropriate delay before merging, would give rise to short 
GRBs at potentially much lower redshift (although see Ripamonti \& Abel 2004 and O'Shea
\& Norman 2007, both of whom find
that fragmentation does not occur in pristine clouds, implying that no
Population III binaries would form). It has been argued that GRBs from such binaries, if they do form, 
would result in negligible detection rates for current detectors, including {\em Swift} 
(Belczynski et al. 2007).

The transition from Population II to Population I star formation begins at $z \sim 7$ 
for the fast or at $z \sim 3$ for the slow metallicity evolution model, continuing 
gradually until the present. Regardless of the metallicity evolution model, Population 
I stars are the dominant contribution to the star formation rate below $z \sim 2$, although 
a small number of Population II stars are also predicted at these lower redshifts. 
The earliest Population I stars begin forming at $z \sim 7$, but this is a negligible
fraction of the total star formation at this redshift, even for the extreme
fast metallicity evolution model (see Fig.~\ref{sfr1}; top panel). Significant formation of 
Population I stars is found for redshifts $z \lesssim 6$ (fast) and $z \lesssim 2.5$ 
(slow), and thus Population I stars are unlikely to be the progenitors of
high-$z$ bursts.

Regardless of metallicity evolution models, Population II stars dominate 
the star formation rate at redshift $6\lesssim z\lesssim10$ (see Fig.~\ref{sfr1}), and are  
the likely progenitors for both GRB 080913 and GRB 090423. Therefore, in the following we
focus exclusively on Population II stars.

\section{Stellar Evolution: Model/Results}

{\em Evolutionary Code.} 
We use the {\tt StarTrack} population synthesis code (Belczynski et al. 2002, 2008a), 
which employs the stellar evolution models of Hurley, Pols, \& Tout (2000), to calculate the 
population of low metallicity stars, with $Z=0.0001$ ($\sim 10^{-2.3} Z_\odot$). 
We evolve $N_{\rm bin}=2 \times 10^6$ 
massive binary systems, with the mass of the primary (the initially more massive) member, 
$M_1$, drawn from a power-law initial mass function 
(IMF) with exponent $-2.7$ in the range $5\mbox{--}150 \,\msun$. The secondary
member's mass, $M_2$, is drawn randomly from a uniform initial mass ratio
distribution ($q=M_2/M_1 \in 0-1$), where we only evolve binaries that have 
secondaries with $M_2>3\,\msun$. Initial orbital separations, $a$, are drawn from a 
uniform random distribution in $\ln a$ (i.e., pdf $\propto 1/a$; Abt 1983), 
and eccentricities are taken from the distribution $\mbox{pdf}=2e$ (e.g., Heggie 1975;
Duquennoy \& Mayor 1991).

We also evolve a population of $N_{\rm sin}=2 \times 10^6$ massive single stars, with 
initial masses chosen as described above for the primary members of the binaries. 
The lower mass limits are chosen such that all stars that can potentially form either
neutron stars or black holes are included in the calculations. 

If the mass range is extended to the hydrogen-burning limit ($\sim 0.08\,\msun$),
one can estimate the total stellar mass that corresponds to our calculation
(i.e., we evolve only the fraction of mass that is formed in massive stars, but 
for the purpose of rate calculations we also estimate the total stellar mass). 
We use a 3-component power law IMF; for stars with masses $0.08\mbox{--}0.5\,\msun$ the
exponent is $-1.3$, for stars in the mass range $0.5\mbox{--}1\,\msun$ we use
$-2.2$, and for masses above $1\,\msun$ we use $-2.7$ (Kroupa, Tout, \& Gilmore
1993; Kroupa \& Weidner 2003).
Note that our population assumes an equal number of single and binary star systems,
corresponding to a binary fraction $f_{\rm bi}=50\%$ (or a binary system membership
probability of 2/3 for any randomly selected star). Spectroscopic studies
demonstrate that binary fractions are significant, especially for massive
stellar populations (e.g., $\gtrsim 70\%$ for the Westerlund-1 cluster (Clark et
al. 2008), or $\gtrsim 50\%$ for NGC6231 (Sana et al. 2008)). 
The total mass in our simulation, in single and binary stars over the entire
mass range, is $M_{\rm sim}=6.1 \times 10^8\,\msun$. 

{\em Short GRBs.} 
We extract from the population synthesis sample the NS-NS and BH-NS binaries with 
delay times shorter than 15 Gyr. The total number of short GRB progenitors is found 
to be $n_{\rm short}=7.6 \times 10^3$ with $4.6 \times 10^3$ NS-NS and 
$3.0 \times 10^3$ BH-NS systems. The delay time, $t_{\rm del}$, is the time 
a given binary takes from the zero age main sequence to form a double compact 
object (evolutionary time, $t_{\rm evol}$) plus the time a double compact object 
takes to merge due to emission of gravitational radiation (merger time, $t_{\rm
merger}$). The resulting delay time distribution is shown in Figure~\ref{del}. 
The median delay time is 0.1 Gyr, with a mean of 1.5 Gyr and a standard 
deviation of 2.9 Gyr. Contrary to conventional wisdom, most mergers are expected 
shortly after the stars form, with only a small fraction (3.7\%) of delay times in
excess of 10 Gyr (qualitatively similar results for Population I stars were
presented in Belczynski et al. 2006). 

{\em Long GRBs.}  We extract black holes (formed in both isolated stars and
in binaries) that {\em (i)} formed through direct collapse (e.g., Fryer \&
Kalogera 2001) and {\em (ii)} formed from a progenitor that at the time of 
core collapse was hydrogen depleted. 
The first condition ensures rapid
accretion ($\sim 0.1-1 \,\msun$ s$^{-1}$) onto a black hole. Popham et
al. (1999) found that at least some of the power sources for GRBs are much more
efficient at higher accretion rates, and Young \& Fryer (2007) argue that
current observations are better fit by these direct-collapse black holes. For
the adopted Population II metallicity, single stars with initial mass $M_{\rm zams} >
23.8 \,\msun$ form black holes through direct collapse. Note that the limiting
mass was $40.0 \,\msun$ for solar metallicity, as obtained in Fryer \& Kalogera
(2001). Here we use different evolutionary models (Hurley et al. 2000)
and we employ stellar winds that scale with metallicity as $\propto Z^{1/2}$, 
and find that stars with lower mass can potentially form black holes directly. 
However, we note that had we used $40.0 \,\msun$ as a limiting mass for direct 
black hole formation, our predicted rates for long GRBs would have  
remained virtually the same since condition {\em (ii)} selects
stars with mass that are above or close to the previous limit (see below). Our limit, which 
takes into account the effect of metallicity and stellar winds, is further 
modified by the presence of a binary companion (from rejuvenation and/or
mass loss in close binaries). 

The second condition allows for a GRB jet to punch 
through the outer layers of a star. A star may lose its H-rich envelope either 
through binary interactions or via stellar winds. Given the employed single star 
models, and the adopted metallicity, stars with $M_{\rm zams} > 35.9 \,\msun$ lose 
their entire H-rich envelopes via stellar winds. Binary interactions, on the 
other hand, may remove the envelope of a star of arbitrary mass. These conditions 
are insufficient to produce a GRB, since there must be just enough angular 
momentum in a collapsing star to form a lasting accretion torus (e.g., MacFadyen 
\& Woosley 1999; Podsiadlowski et al. 2004). As angular momentum transport is not 
taken into account in this study, our predictions only indicate potential long GRB 
progenitors, i.e. most likely only a (small) fraction of the events that 
satisfy {\em (i)} and {\em (ii)} will actually produce long GRBs. 

The total number of long GRB progenitors is found to be $n_{\rm long}=225 \times 10^3$ 
with $103 \times 10^3$ and $122 \times 10^3$ direct black holes (DBHs) from
H-depleted progenitors from single and binary stars, respectively. The delay
time distribution is displayed in Figure~\ref{del}. This represents the
evolutionary time ($t_{\rm evol}$) of a given star to form a black hole. The delay 
times for long GRB progenitors are very short, $\sim 3.5\mbox{--}12$ Myr (as black 
holes form from massive and short-lived stars), with the median of the 
distribution at 5.1 Myr, a mean at 5.3 Myr, and a standard deviation of 1.1 Myr.
These events thus trail the star formation rate, SFR$(z)$, with very little delay, so 
that essentially one can assume that Rate(DBH) $\propto$ SFR until very large redshifts.

\section{Cosmology: Model}

We adopt a flat cosmology with $H_0=70.0\mbox{ km s}^{-1}\mbox{Mpc}^{-1}$, 
$\Omega_m=0.3$, and $\Omega_\Lambda=0.7$ (and thus $\Omega_{\rm k}=0$). 
Small ($\lesssim 10\%$) changes in the cosmological parameters 
leave our results essentially unaltered.

For clarity in what follows, we briefly review the relevant cosmological
expressions. The relationship between redshift and time is given by
\begin{equation}
t(z)=t_{\rm H} \int_{z}^{\infty} {dz' \over (1+z') E(z')},
\label{eq8}
\end{equation}
where $t_{\rm H}=1/H_0=14$ Gyr is the Hubble time (e.g., Hogg 2000), and $E(z)=
\sqrt{\Omega_{\rm M}
(1+z)^3+\Omega_{\rm k}(1+z)^2+\Omega_\Lambda)}$.
The resulting age of the
Universe is $t(0)=13.47$ Gyr. It is to be noted that this is the restframe time,
constituting what is measured by the wristwatches of local observers at any time
and place in the Universe, and thus is the appropriate quantity to use when
discussing local physical processes. Observed event rates from sources distributed
over cosmic time must be corrected for time dilation, $R(0) = R(z)/(1+z)$.

The comoving volume element, $dV/dz$, for a given solid angle, $d\Omega$, is given by
\begin{equation}
{dV \over dz}(z) = {c \over H_0} {D_{\rm c}{}^2 \over
E(z)}\,d\Omega,
\label{eq9}
\end{equation}
where $c$ is the speed of light in vacuum, and where the comoving distance, $D_{\rm c}$, 
is given by
\begin{equation}
D_{\rm c}(z) = {c \over H_0} \int_{0}^{z} {dz' \over E(z')}.
\label{eq10}
\end{equation} 

There are a number of steps necessary to calculate the expected GRB event rates. 
To summarize: for each bin in time, we estimate the total star formation rate. We
then calculate the fraction of these stars which are Population II by incorporating 
the metallicity distribution of the galaxies at the given time. We then employ 
population synthesis results to determine the fraction of Population II stars that 
are progenitors of GRBs. 
With these results, we are able to apply the appropriate delay times and transform 
the progenitor formation rate in each bin in time to an actual GRB event rate, 
distributed over later times. We repeat this procedure over all bins in time, 
which results in a total GRB event rate as a function of time. We now go through 
each of these steps in detail.

We bin time over the range $0.13<t<13.47\mbox{ Gyr}$ ($0<z<25$), with a constant width
of $dt=10$ Myr. 
For each bin we obtain the star formation rate using Equation~\ref{eq1}, evaluated at
the center of the bin. We then generate a Monte Carlo sample of $10^4$ galaxies
in the bin, distributed according to Equation~\ref{eq2}, with a total mass in 
galaxies given by $M_{\rm gal,tot}$. 
For each galaxy we estimate its average metallicity from Equation~\ref{eq3} and, 
based on our population criterion (Eq.~\ref{eq7}), we determine the number of galaxies 
containing Population II stars. The total mass of galaxies with Population II stars 
is denoted $M_{\rm gal,II}$. The fraction of mass that has formed Population II stars in 
a given bin is thus given by
\begin{equation}
F_{\rm pop,II}={M_{\rm gal,II} \over M_{\rm gal,tot}}, 
\label{eq15}
\end{equation}
since we have assumed that each galaxy hosts only one stellar population
(determined by its average metallicity; see \S\,2).

We measure rate densities in units of ${\rm yr}^{-1}\ 
{\rm Gpc}^{-3}$, in comoving units (time and space). In our scheme a delay for a 
given event constitutes a shift in time of the relevant fraction of the rate density 
(see Eq.~\ref{eq15b}).
The population synthesis results yield a list of Population II 
GRB progenitors, and their corresponding delay times, and we use all of them in each 
time bin. Thus, for each bin we have a total of $n_{\rm short}=7.6 \times 10^3$ and 
$n_{\rm long}=225 \times 10^3$ progenitors for short and long GRBs, respectively. 
Each progenitor in our list now represents a fraction of the total star
formation rate density in the given bin that generates GRBs. This fraction is 
given by: 
\begin{equation}
f_{\rm sim}={F_{\rm pop,II} \over M_{\rm sim}}{\rm sfr}(t),
\label{eq15b}
\end{equation}
where $M_{\rm sim}$ is the total stellar mass corresponding to our population
synthesis simulations (see \S\,3), and $f_{\rm sim}$ has units of 
${\rm yr}^{-1}\ {\rm Gpc}^{-3}$.

For each progenitor in the list we choose a random starting time, $t_0$, within 
the given time bin, and propagate the progenitor formation rate density,
$f_{\rm sim}$, forward in time to
\begin{equation}
t_{\rm new}=t_0 + t_{\rm del}.
\label{eq16}
\end{equation}
The delay time, $t_{\rm del}$, marks the time elapsed from the formation of the 
star to the potential GRB event, as described in \S\,3, and   
$t_{\rm new}$ is the time at which a given GRB event
actually occurs. We thus convert the formation rate density of progenitors into GRB event
rate densities. By repeating the above series of steps for each time bin, we arrive at 
a total comoving GRB event rate density, $n_{\rm rest}(t)$.

\section{Intrinsic rates} 

In this section we calculate the intrinsic rate of short and long GRBs from
Population II progenitors. In the first subsection we discuss correction factors
due to model uncertainties, while in the second subsection we provide estimates
of the intrinsic GRB rates.

\subsection{Correction factors: population synthesis}

Both the direct black hole formation and double compact object merger rates 
obtained from population synthesis are only first-order approximations for long 
and short duration GRBs, respectively.
For example, the only constraints placed on our long-duration GRB progenitors were 
the assumption that the black hole formed from direct collapse and that the star, 
at collapse, was a He star. Observations of nearby GRB/SN associations suggest 
that the progenitor must lose most of its Helium, because, to date, every supernova 
observed associated with a GRB is a type Ic supernova (see Fryer et al. 2007 for a 
recent review). We have not placed additional constraints requiring the loss of the He
envelope. In addition, we have not placed constraints on the angular momentum
profile, which is likely to play an important role in GRB formation. 

The type Ic constraint poses a problem for many of the current
progenitor scenarios and, hence, makes it difficult to predict a reliable
rate from population synthesis studies alone. Ignoring this constraint, we can
make only rudimentary estimates of the GRB event rates. With the
rotation rates suggested by Yoon et al. (2006), roughly 10\% of
the collapsing stars in our simulations form GRBs. Given our rate estimates
for both long and short bursts, this indicates that the odds are roughly 
10 to 1 that GRB 080913 and 090423 were long-duration bursts.
However, Podsiadlowski et al.~(2004) have argued that only 1\% of all black hole
systems would form GRBs. Although these authors mention that this fraction could
potentially increase at low metallicity, if we keep their 1\% figure 
we find a long-duration GRB rate 100 times smaller than the one we calculated for
the direct black hole formation channel. Thus, the likely rate for long-duration GRBs 
is reduced by a factor of $\epsilon_{\rm syn,long} \sim\mbox{0.01--0.1}$.  

Our rate for short GRBs is also an over-estimate. We include all BH-NS mergers, 
while neglecting system configuration (inclination/mass ratio) and black hole 
spin (only 1--40\% percent of mergers are expected to form BH torus configurations
and thus potentially result in a GRB: Belczynski et al. 2008b). If GRBs indeed require
the formation of a black hole (e.g., Janka \& Ruffert 1996; Oechslin \& Janka 
2006; Lee \& Ramirez-Ruiz 2007), the fraction of NS-NS mergers is limited by 
the maximum neutron star mass. Most known NS-NS mergers will form central compact 
objects with masses $M_{\rm c} \lesssim 2.5 \,\msun$ (Belczynski et al. 2008c). The maximum
neutron star mass must be less than this value, or these mergers will not form black 
hole accretion disks. However, the requirement of a black hole is a weak constraint as 
neutron star accretion disks as well as proto-magnetars may also produce short GRBs
(Usov 1992; Kluzniak \& Ruderman 1998; Dai et al. 2006; Metzger, Quataert \& Thompson 
2007). The low metallicity population of double compact objects consists of $\sim 60\%$ 
of NS-NS binaries and $\sim 40\%$ of BH-NS systems (see \S\,3). Thus the reduction due
to the required configuration of BH-NS systems does not lead to a significant
decrease of short GRB rates. The likely rate for short-duration GRBs 
is thus reduced by approximately $\epsilon_{\rm syn,short} \sim\mbox{0.6--0.8}$.

Additionally, the formation rates of double compact objects as calculated in
population synthesis are subject to uncertainties that are associated with some 
poorly understood evolutionary processes that are important in the evolution of
massive binaries (e.g., winds, mass transfer episodes, natal kicks). Belczynski 
et al. (2002) presented a comprehensive parameter study of double compact object
formation rates. If the most unrealistic and unphysical models are omitted,
it was shown that NS-NS and BH-NS merger rates vary at most by factor of
$\sim 6$ (up and down) from the reference model (best guess for evolutionary 
parameters) value. We use this as an estimate of the population synthesis uncertainty
in the rates and thus renormalize correction factors to $\epsilon_{\rm syn,short}
\sim\mbox{0.1--4.8}$.

In summary, the population synthesis or model related correction factors are 
\begin{equation}
  \epsilon_{\rm syn,long}=0.01\mbox{--}0.1 
  \label{eq20a}
\end{equation}
\begin{equation}
  \epsilon_{\rm syn,short}=0.1\mbox{--}4.8  .
  \label{eq20b}
\end{equation}

\subsection{Intrinsic rate estimate}
Our estimate of the intrinsic GRB event rate density in the rest frame is given by
\begin{equation}
{\cal N}_{\rm rest}(t) = \epsilon_{\rm syn}\ n_{\rm rest}(t)\ {\rm yr}^{-1}\,{\rm Gpc}^{-3},
\label{eq18}
\end{equation}
where $\epsilon_{\rm syn}$ was introduced in the previous subsection, and $n_{\rm
rest}$ was introduced in \S\,4.
The rate density of GRBs that would be observed is appropriately time dilated:
${\cal N}_{\rm obs}(t) = {\cal N}_{\rm rest}(t)/(1+z(t))$.
In a slightly confusing accident of notation, the units of both 
of these quantities is ${\rm yr}^{-1}\ {\rm Gpc}^{-3}$, where in one case years
are measured in the rest frame of the GRB population, and in the other case years 
are measured by a clock at the local observatory.

It is to be noted that some events have such long delay times, or are associated 
with such recent star formation, that the GRB events will happen in our future 
($t_{\rm new}>13.47$ Gyr), and thus are not included in the event rate
density estimates. We note that this affects only short GRB progenitors, 
as delay times for long GRB progenitors are very short (see Fig.~\ref{del}).

We are also interested in the intrinsic GRB event rate, in addition to the event 
rate density. We present intrinsic event rates in the observer frame. To determine 
the rate to some specified redshift, $z$, we integrate the GRB event rate density 
(modified by time dilation) over the comoving volume:
\begin{equation}
N(<z)= 4 \pi \epsilon_{\rm syn} \int_{0}^{z} {n_{\rm rest} \over 1+z'} {dV \over
 dz'}dz'\ \ {\rm yr}^{-1},
\label{eq19}
\end{equation}
where we have integrated over the entire sky ($\int\! d\Omega = 4 \pi$). It is
important to note that, although the above integral is in redshift, it is
integrating out the volume (given by Eq.~\ref{eq9}), and not the time: the
original units of rate density (${\rm yr}^{-1}\ {\rm Gpc}^{-3}$) have been converted 
into a rate (${\rm yr}^{-1}$) and not into a density (${\rm Gpc}^{-3}$). $N(<z)$ 
represents the rate in the observer frame of all GRB events (bursts per year)
integrating to a limiting redshift of $z$. Since $N$ denotes the intrinsic
rate, we have not yet applied observational selection effects due to such things as 
beaming, brightness, or instrumental response (these are calculated in \S\,6.1).

The population synthesis code naturally distinguishes between long and short 
GRBs, and we carry this identification through our calculations, quoting results 
for both long ($N_{\rm long}$) and short ($N_{\rm short}$) GRBs. The resulting 
intrinsic GRB rate densities and event rates are shown in
Figures~\ref{rate2},~\ref{rate3}, and~\ref{rate4}. As expected for long GRBs,
these rates closely 
track the star formation history of Population II stars: the majority ($90\%$) 
of events are found at redshifts $z \sim 3\mbox{--}15$ with a peak at $z \sim 7$ 
(fast metallicity evolution model), or at $z \sim 2\mbox{--}11$ with a peak at $z \sim 3$ 
(slow model). Short Population II GRBs start appearing slightly later than long 
Population II GRBs, and their distribution extends to lower redshift ($z \sim 0$). 
However, the peak of the short GRB rate is near the long burst peak, at $z \sim 7$ 
(fast) and $z \sim 3$ (slow), due to the large abundance of very short delays. We 
note that even a delay as short as 0.1 Gyr is sufficient to result in a many kpc 
distance between the site of star formation in the host galaxy and the eventual merger 
location in the halo (e.g., Fryer, Woosley, \& Hartmann 1999; Bloom, Sigurdsson, \& 
Pols 1999; Belczynski et al. 2006). The short GRB progenitors considered here 
(NS-NS/BH-NS mergers) have a tail of very long delay times (median 0.1 Gyr, and 
standard deviation 2.9 Gyr), resulting in substantial intrinsic rates of Population 
II mergers at low redshift. 

We note that in the redshift range $6<z<8$, the intrinsic
rates for long and short GRBs are comparable (see Fig.~\ref{rate4}). At
redshift $z=6.7$, the rates are $N_{\rm long} = 10 \times 10^3\mbox{--}153 \times
10^3$ yr$^{-1}$ and $N_{\rm short} = 3 \times 10^3\mbox{--}141 \times 10^3$ yr$^{-1}$ 
per unit redshift, in the observer frame. At redshift $z=8.1$ these rates 
become $N_{\rm long} = 9 \times 10^3\mbox{--}91 \times 10^3$ yr$^{-1}$ and 
$N_{\rm short} = 2 \times 10^3\mbox{--}84 \times 10^3$ yr$^{-1}$ (the ranges 
encompass both of our metallicity evolution models). The above similarity of 
rates arises for two reasons: first, the population synthesis rates of short 
($\epsilon_{\rm syn,short} n_{\rm short} = 0.8 \times 10^3\mbox{--}36.5 \times 10^3$) 
and long 
($\epsilon_{\rm syn,long}  n_{\rm long}  = 2.3 \times 10^3\mbox{--}22.5 \times 10^3$) 
GRB progenitors are comparable; second, Population II star formation rates for the two 
metallicity evolution models are almost the same 
($\sim 0.1\,\mpy$ Mpc$^{-3}$; see Fig.~\ref{sfr1}) at these redshifts. 

For comparison, we have also calculated the rate of supernovae utilizing the 
same procedure presented above. The cumulative rate of core collapse supernovae 
(Type II and Ib/c) is estimated to be $\sim 8$ s$^{-1}$ (integrated to redshift 
$z=10$), which is comparable to the rates from recent empirical estimates of 
SFR$(z)$, e.g. Hopkins and Beacom~(2006).

\section{Observed rates}

In this section we predict the {\em Swift} detection rate of short and long GRBs from
Population II progenitors. In the first subsection we discuss observational
selection effects, while in the second subsection we provide estimates
of the observed GRB rates.

\subsection{Correction factors: observational selection effects}

Not all high-redshift GRBs will necessarily be observable. There are
many factors which impact whether or not a given GRB will be observed, including
the lightcurve and spectral energy distribution of the burst (both of them
redshifted), as well as the spectral and time sensitivity of the detector. A
precise calculation of the selection function of GRBs is well beyond the scope
of this paper. The GRB population is tremendously heterogeneous, and the
triggering algorithms of an instrument such as the BAT on {\em Swift}\/ are highly
complex (Band~2003,~2006; Fenimore et al.~2004). In what follows we make a number
of simplifications, and derive an approximation for the observed rate of short and 
long GRBs. 

We assume that {\em all}\/ GRBs have the same distribution of average isotropic 
luminosity, $L_{\rm iso}$, given by a Gaussian in $\log(L_{\rm iso})$, centered on 
$L_{\rm iso}=10^{50}$ erg s$^{-1}$, with $1\sigma=1\ \mbox{dex}$. We assume all
short GRBs have a duration of $t_{\rm 90}=0.3\mbox{ s}$, while all long GRBs
last $t_{\rm 90}=30\mbox{ s}$. The distributions of total energy (integrated flux;
calculated as ${\cal E}=L_{\rm iso} t_{\rm 90}$) for short and long bursts are 
presented in Figure~\ref{sed}. 

We approximate the {\em Swift}\/ GRB trigger (for the BAT instrument) as a 
fluence\footnote{Our fluence threshold can easily be rephrased as a trigger 
on peak flux/luminosity, with all of the units being converted from [erg] to 
[erg s$^{-1}$]. The results are identical. In addition, we are calculating
the fluence from an isotropic luminosity, but the results are unchanged if
this is taken to be a beamed luminosity.} threshold. 
We take the long GRB trigger as 
${\bar{\cal E}}_{\rm long}=3\times10^{50}\ \mbox{erg}$ for a long burst at 
$z=1$ (see, e.g., Fig.~2 of Kistler et al.~2008 and Fig.~2 of Butler et al.~2007). 
In other words, a long GRB at $z=1$ must emit a total isotropic-equivalent
energy satisfying 
${\cal E} \geq 3 \times 10^{50}\ \mbox{erg}$ to be observable. The corresponding 
short GRB threshold is given by ${\bar{\cal E}}_{\rm short}=10^{50}\ \mbox{erg}$ 
at $z=1$. The difference between the thresholds is chosen to mimic some 
of the complexity in the BAT triggers, including integration and readout times, 
as well as sensitivity to the differing spectral energy distributions (SEDs) of 
long and short bursts (Fenimore et al.~2004; Band~2003,~2006). For a burst at 
redshift $z$, we draw an energy from the relevant distribution shown in 
Figure~\ref{sed}, and then calculate whether the fluence of the burst satisfies 
our trigger threshold criterion:
\begin{equation}
{{\cal E}}\geq\left({D_L(z)/D_L(z=1)}\right)^2 \bar{\cal E},
\end{equation}
where $D_L(z)$ is the luminosity distance out to redshift $z$.
Only those bursts which satisfy this inequality are taken to be observed. The
resulting fraction of long and short bursts observed with {\it Swift}, $\epsilon_{\rm
SED}$, as a function of redshift, is plotted in Figure~\ref{sed}. 

Another factor we have not yet considered is beaming, and the resulting visibility of GRBs. 
In general short GRB jets appear to be wider, with opening angles of 
$\Theta \gtrsim 10^\circ$ (e.g., Fox et al. 2005; Berger 2007; Metzger, Piro, \& 
Quataert 2008) as compared with long GRBs, $\Theta \sim 5^\circ$ (e.g., Frail et
al. 2001; Bloom, Frail, \& Kulkarni 2003; Grupe et al. 2006; Soderberg et al. 2006). 
This would imply that long GRBs are $\sim 4$ times less likely to be detected, leading
to further reduction of the observed rates by factors of 
\begin{equation}
  \epsilon_{\rm beam,long}=0.002
  \label{beam1}
\end{equation}
\begin{equation}
  \epsilon_{\rm beam,short}=0.008
  \label{beam2}
\end{equation}
for long and short GRBs, respectively. 

\subsection{Observed rate estimate}

We now provide a prediction for the {\em Swift} detection rate for 
Population II GRBs. We express rates per unit redshift to
emphasize the redshift range relevant for GRB 080913 and GRB 090423. 
The predicted detection rate is given by:
\begin{equation}
{dN_{\mbox{\tiny Swift}} \over dz} = f_{\rm Swift}\ \epsilon_{\rm SED}\ \epsilon_{\rm beam}\ 
                         {dN \over dz}\ {\rm yr}^{-1},
\end{equation}
where $N$ is the intrinsic rate of GRBs (see \S\,5.2), $\epsilon_{\rm beam}$
is a correction factor for GRB beaming (see \S\,6.1), $\epsilon_{\rm SED}$ is the fraction of GRBs 
above the {\em Swift}\/ detection threshold (see Fig.~\ref{sed} and \S\,6.1), and $f_{\rm Swift}=1.4/(4
\pi)=0.1$ represents the {\em Swift}\/ sky coverage (Gehrels et al. 2004; Barthelmy et
al. 2005). We plot our prediction for the observed rate of both short and long
GRBs in {\em Swift}\/ in Figure~\ref{fig:swift_rate}.
In the redshift range $z=6.2\mbox{--}7.2$ centered on $z=6.7$ (GRB 080913), we
expect {\em Swift}\/ GRB detection rates of:
\begin{eqnarray}
N_{\mbox{\tiny Swift},{\rm long}} &\sim & 
             0.33\mbox{--}5.0\mbox{ yr}^{-1}\mbox{ at $z=6.2\mbox{--}7.2$}\\
N_{\mbox{\tiny Swift},{\rm short}} &\sim & 
             0.01\mbox{--}0.70\mbox{ yr}^{-1}\mbox{ at $z=6.2\mbox{--}7.2$}.
\label{eq21}
\end{eqnarray}
For the redshift range $z=7.6\mbox{--}8.6$ centered on
$z=8.1$ (GRB 090423), the predicted {\em Swift}\/
detection rates become:
\begin{eqnarray}
N_{\mbox{\tiny\it Swift},{\rm long}} &\sim  & 
         0.21\mbox{--}2.2\mbox{ yr}^{-1}\mbox{ at $z=7.6\mbox{--}8.6$}\\
N_{\mbox{\tiny\it Swift},{\rm short}} &\sim & 
         0.004\mbox{--}0.24\mbox{ yr}^{-1}\mbox{ at $z=7.6\mbox{--}8.6$}.
\label{eq21b}
\end{eqnarray}

These results suggest that GRB 080913 and GRB 090423 are most likely associated
with the deaths of massive stars (long GRBs), rather than double compact object 
mergers (short GRBs). At redshifts $z=6.7$ (GRB 080913) and 
$z=8.1$ (GRB 090423) long GRBs are expected to be $\sim 10$ times more frequent 
than short GRBs in the {\em Swift}\/ sample. We emphasize
that we have taken an optimistic value for the short burst observational 
threshold; the true short GRB rate may be up to an order of magnitude smaller
(see \S\,6.1 for details). 
Our results thus represent an upper-limit to the short GRB fraction at these 
high redshifts. Although the intrinsic event rates for
collapsars and double compact object mergers are similar at these high
redshifts, the observational difference is due to the very distinct
observational selection effects between these two event types. We note that our
results are consistent with Zhang et al.~(2009), who classified both bursts
as collapsars based upon their observational properties. 
We have ignored recent evidence that the long GRB population evolves with
redshift, increasing the number of observed bursts at high redshift. In this
sense, our results can be considered a lower limit to the long GRB rate
(although we note that the corresponding evolution of short bursts is
currently poorly constrained). Further details of the nature of GRB 
intrinsic properties and observational selection effects can be found for
example in Guetta, Piran \& Waxman 2005;  Guetta \& Piran 2006; Berger 2007; 
Kistler et al. 2008; Salvaterra et al. 2009 and Virgili, Liang \& Zhang 2009.

\section{Summary}

We have calculated the evolution of stars at high redshift, exploring possible
progenitors for GRB 080913 (at $z=6.7$) and GRB 090423 (at $z=8.1$). We find
that in the redshift range $6\lesssim z\lesssim10$ the majority of stars are
Population II (low metallicity). By this time Population III stars (metal free)
have already finished their evolution, and Population I stars (metal rich) are
only just beginning to form. We have adopted two extreme models for the
metallicity evolution history, which most likely bracket the true history; our
final results are insensitive to the choice of model. We find that the
progenitors of GRB 080913 and GRB 090423 are likely to have been Population II
stars.

\begin{figure}
\includegraphics[width=1.0\columnwidth]{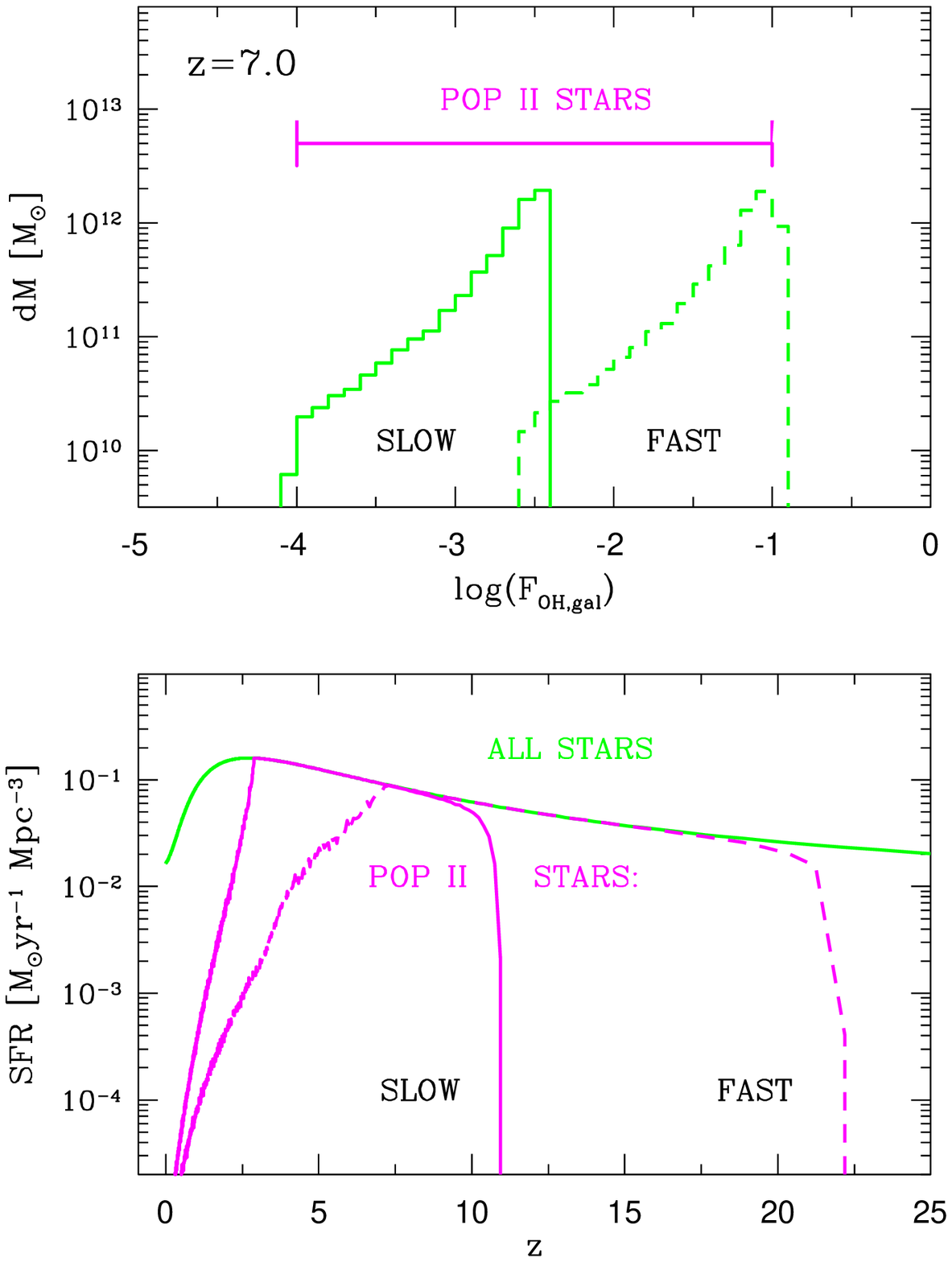}
\caption{
{\em Bottom panel:} Star formation rate history. We have adopted an extinction corrected 
SFR model from Strolger et al. (2004): all stars. We also show the rate we have obtained 
for just Population II stars (see \S\,2): the population is rather different for
slow (solid) and fast (dashed line) metallicity evolution model.
{\em Top panel:} Metallicity distribution of stars at redshift $z=7.0$.
The results are shown for the two different adopted metallicity evolution histories: 
fast (Young \& Fryer 2007) and slow (Pei et al. 1999). 
We mark our definition for Population II stars: it is clearly seen that
independent of metallicity evolution majority of the stars belong to Population
II at redshift that is typical for stars that may become (after appropriate
delay) progenitors of GRB 080913. 
}
\label{sfr1}
\end{figure}

\begin{figure}
\includegraphics[width=1.0\columnwidth]{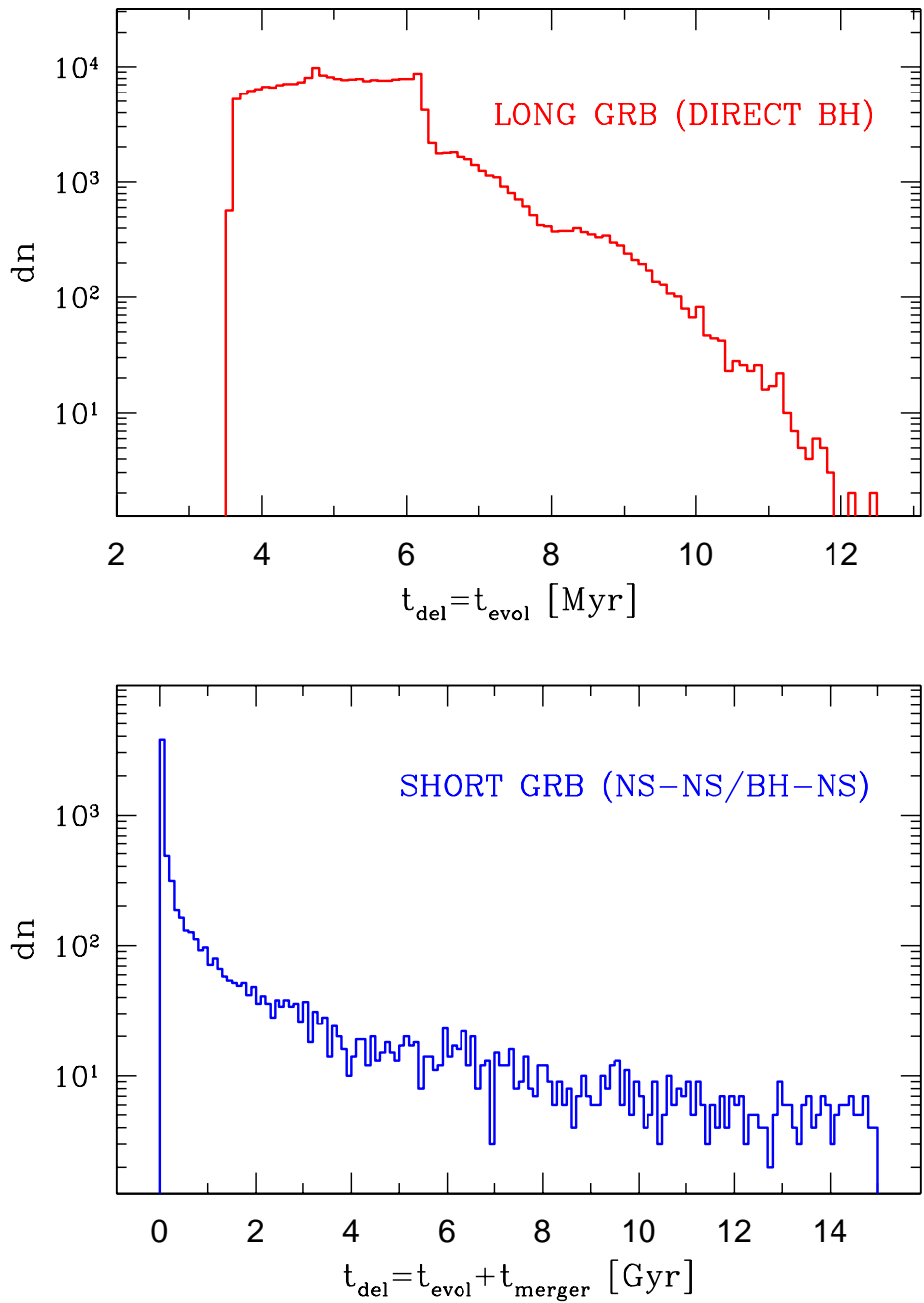}
\caption{
Top panel: Delay times for long GRB progenitors: direct black holes that are formed
out of H-depleted stars (both single and binary). Note the very
short delay times of $\lesssim 6$ Myr. Delay is evolutionary time a star
takes from the formation to a collapse. 
Bottom panel: Delay times for short GRB progenitors: NS-NS and BH-NS mergers.
Note that these events have significantly longer delay times than for
the long GRB progenitors, with a median of $0.1$ Gyr. Delay time is 
evolutionary time plus merger time.}
\label{del}
\end{figure}

\begin{figure}
\includegraphics[width=1.0\columnwidth]{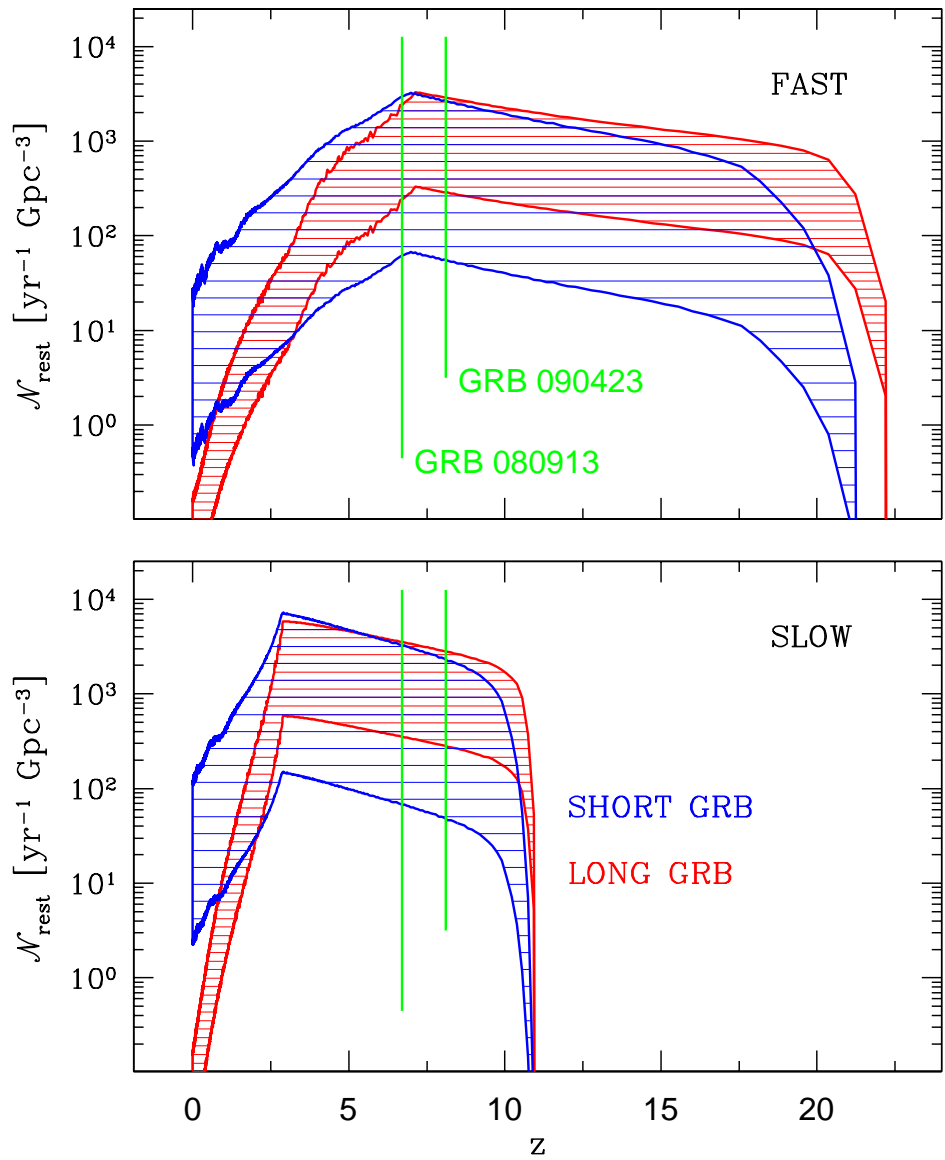}
\caption{
Short and long GRB {\em intrinsic}\/ event rate densities (in the restframe) originating 
from Population II stars for slow and fast metallicity evolution. These are 
expressed per year, per cubic comoving Gpc. 
}
\label{rate2}
\end{figure}

\begin{figure}
\includegraphics[width=1.0\columnwidth]{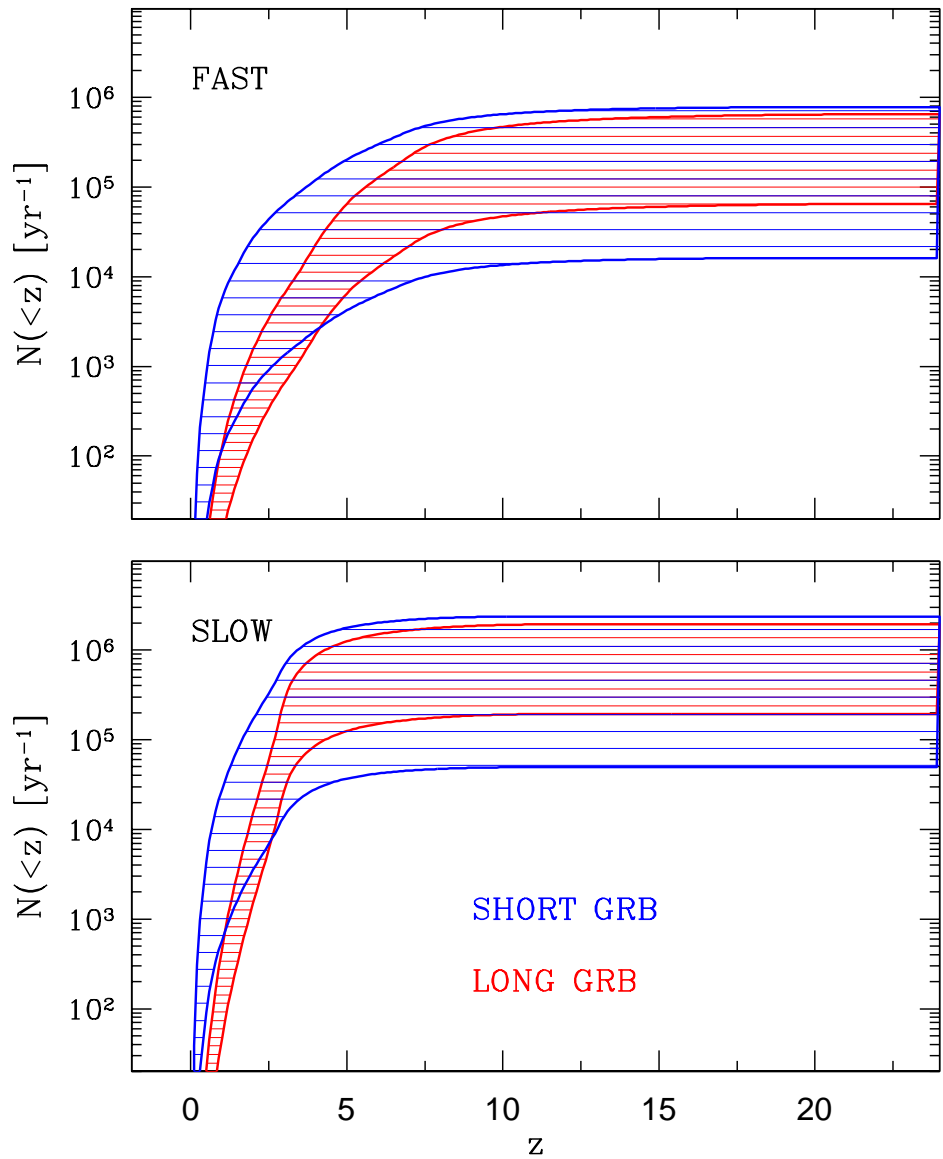}
\caption{
{\em Intrinsic}\/ event rate of short and long GRBs that originate from 
Population II stars, per year (in the observer frame), as a function 
of the depth of the survey in redshift, $z$. The rates are shown for 
slow and fast metallicity evolution models. 
}
\label{rate3}                                  
\end{figure}

\begin{figure}
\includegraphics[width=1.0\columnwidth]{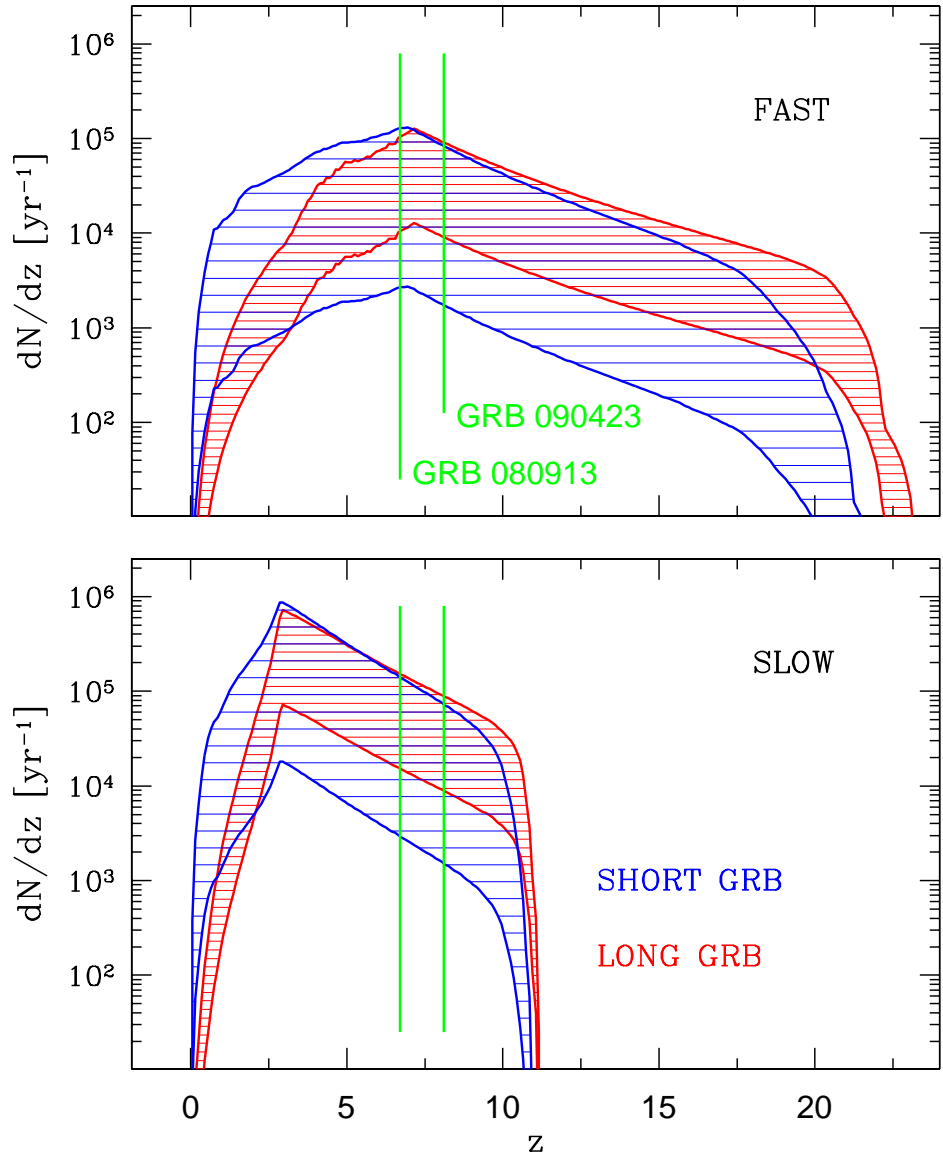}
\caption{ 
{\em Intrinsic}\/ event rate of short and long GRBs that originate from
Population II stars, per year (in the observer frame) presented per
unit redshift. These plots show the derivative of curves presented in
Figure~\ref{rate3}, highlighting the redshift range in which GRB events 
from Population II stars contribute.
}
\label{rate4}                                  
\end{figure}

\begin{figure}
\includegraphics[width=1.0\columnwidth]{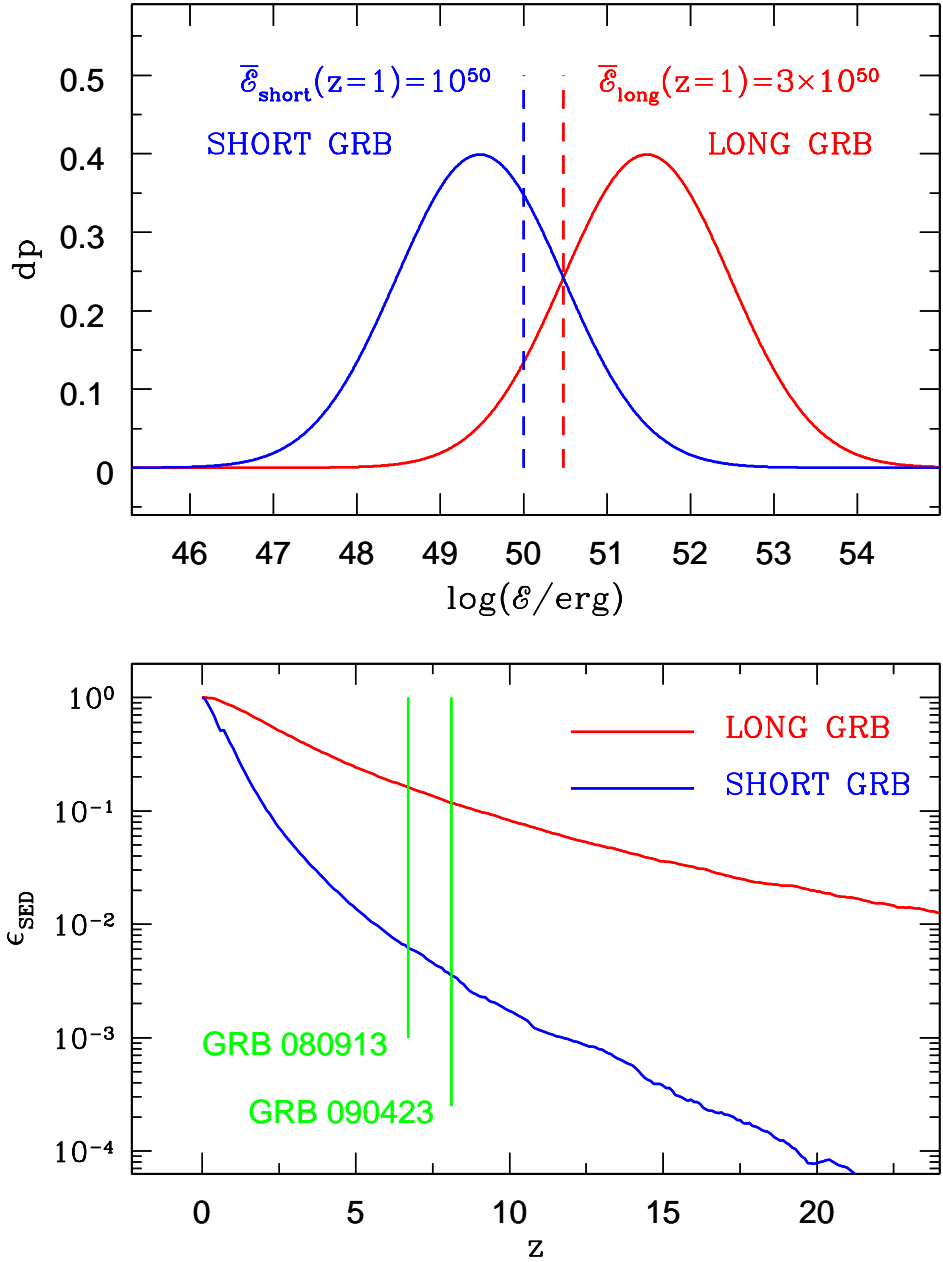}
\caption{ 
{\em Top panel:} Distribution of total energy (${\cal E} = L_{\rm iso} t_{\rm 90}$) 
for short and long GRBs. Short and long GRBs are assumed to have the same
distribution of average isotropic luminosity, $L_{\rm iso}$, described by a
Gaussian in $\log(L_{\rm iso})$ with 
mean at $\log(L_{\rm iso}/[{\rm erg\ s}^{-1}])=50$ and $\sigma =1.0$.
A single typical duration time was assumed for all short ($t_{\rm 90}=0.3$s) 
and long ($t_{\rm 90}=30$s) GRBs. 
The adopted {\em Swift}\/ detection thresholds (at $z=1.0$) are also shown;
$\bar{\cal E} = 10^{50},\ 3\times 10^{50}$erg for short and long GRBs, 
respectively. See the text for details.
{\em Bottom panel:} The fraction of GRBs above the {\em Swift}\/ 
detection limit,  $\epsilon_{SED}$, as a function of redshift.
Note that at redshift $z=6.7$ (GRB 080913) the fraction of detectable long GRBs 
is $\epsilon_{\rm SED,long}=0.162$, while for short events it is only 
$\epsilon_{\rm SED,short}=0.006$. At redshift $z=8.1$ (GRB 090423) the 
difference in detection probability is even larger: 
$\epsilon_{\rm SED,long}=0.118$ and $\epsilon_{\rm SED,short}=0.004$.
}
\label{sed}    
\end{figure}

\begin{figure}
\includegraphics[width=1.0\columnwidth]{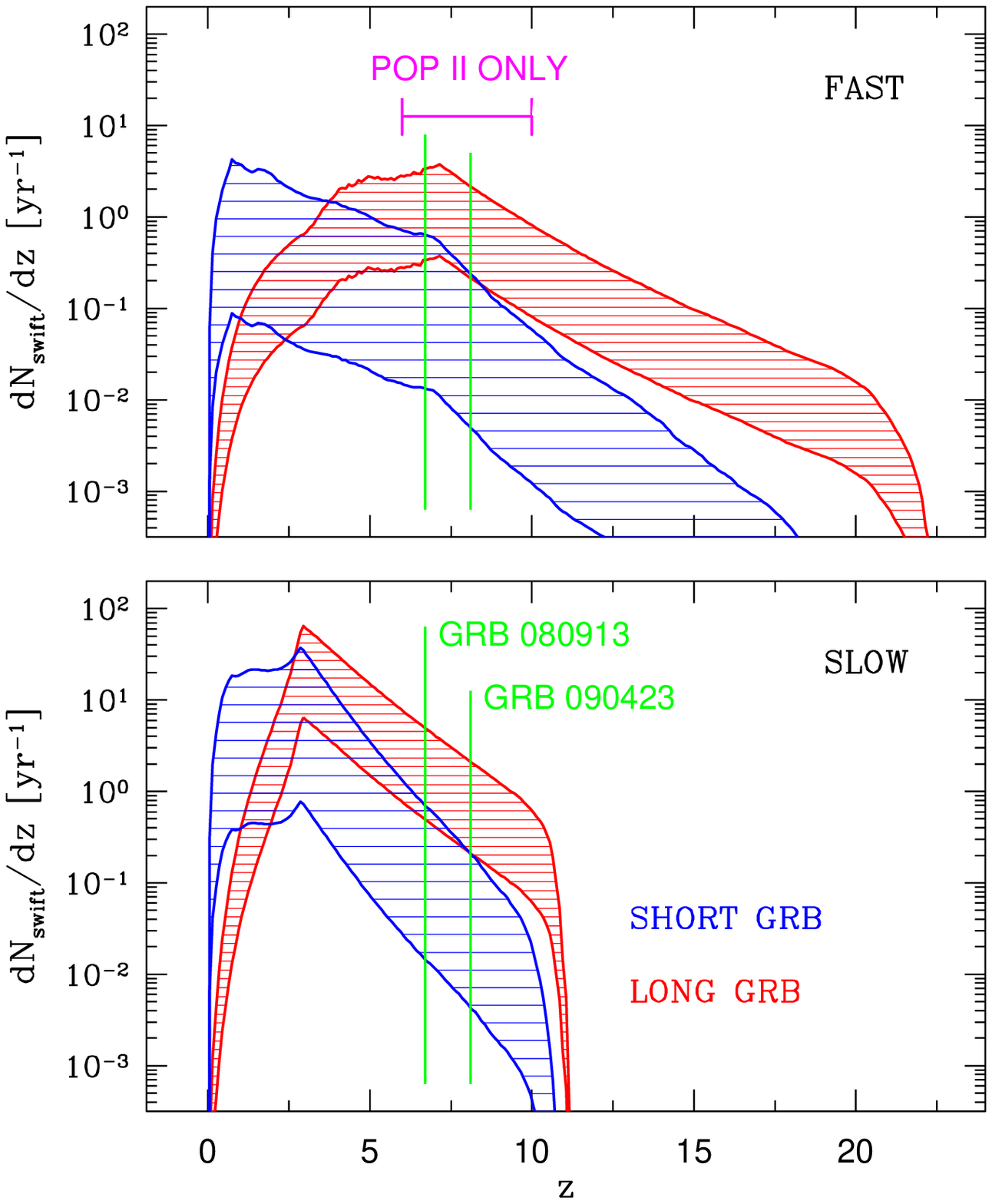}
\caption{
Predicted {\em Swift}\/ {\em detection}\/ rates of short (NS-NS/BH-NS mergers) 
and long (collapsars) GRBs. Rates are for GRBs originating {\em exclusively} 
from Population II stars, and hence can only be compared to {\em Swift}\/
data in the redshift range in which Population II stars are the
dominant stellar population: $6\lesssim z \lesssim10$ (as marked on the plot).
For low redshifts ($z \lesssim 6$) there is a significant contribution to
GRB rates from Population I stars (not shown here), and for very high
redshifts ($z \gtrsim 10$) there may be a contribution from GRBs originating from
Population III stars (not shown here).
Note that GRB 080913 and GRB 090423 are both found in the redshift range in which
the majority of GRB events are predicted to originate from Population II
stars. Furthermore, our results favor the collapsar origin for these GRBs.
The two panels show the rates for a fast (early metal mixing into stars; top
panel) and a slow (late mixing; bottom panel) metallicity evolution model. In
the redshift range of interest ($6\lesssim z\lesssim10$), the rates of the two
extreme models are similar, indicating that our conclusions are robust: GRB
080913 and GRB 090423 have Population II progenitors, and are both long bursts
resulting from collapsars.
}
\label{fig:swift_rate}    
\end{figure}

For Population II stars we subsequently calculated the rates of double compact object
mergers (NS-NS and BH-NS) and the rates of collapsars (deaths of massive
stars). The former are thought to be associated with short GRBs, while the
latter are believed to be responsible for long GRBs. We
find that the intrinsic rates for short GRBs (mergers) and long GRBs
(collapsars) are comparable at high redshift (see Fig.~\ref{rate4}). However,
the observational selection effects (beaming, differing intrinsic fluence, and
instrumental response) make long bursts more likely to be seen by a satellite
like {\em Swift}.  Our main result is shown in Figure~\ref{fig:swift_rate},
where we plot our predicted {\em Swift}\/ detection rates for both short and
long GRBs.  On average, the detection rates for 
long GRBs are $10$ times higher than the rates for short GRBs. At the redshifts
of GRB 080913 and GRB 090423, the rates are $\sim 1$ yr$^{-1}$ and $\sim 0.1$
yr$^{-1}$ per unit redshift, for long and short GRBs, respectively. We emphasize that our
calculations only consider Population II stars, as these stars dominate the star
formation rate at redshift $6<\lesssim z\lesssim10$ (see Fig.~\ref{sfr1}), and
are thus appropriate for the two highest known redshift GRBs. In future work we
will include all generations of stars.

We conclude that both GRB 080913 and GRB 090423 are most likely to have been
long bursts, resulting from the deaths of massive stars (collapsars) from
Population II progenitors. With the observation of GRB 080913 and 090423,
gamma-ray bursts have entered a unique high-redshift regime. As future data
fills out the high-$z$ tail of GRBs, these systems will become one of the most
powerful probes of the star formation, stellar death, and chemical enrichment of
our Universe.

\acknowledgements
We thank the anonymous referee for encouraging us to incorporate observational
selection effects into our analysis. KB acknowledges support from KBN grant 
N N203 302835.

\clearpage

\clearpage

\clearpage

\clearpage

\clearpage

\clearpage

\clearpage


\begin{references}

\reference{} Abt, H.\ A.\ 1983, ARA\&A, 21, 343

\reference{} Amati, L.\ 2006, \mnras, 372, 233 

\reference{} Band, D.L.\ 2003, \apj, 588, 945

\reference{} Band, D.L.\ 2006, \apj, 644, 378

\reference{} Barthelmy, S. D., et al.\ 2005,  Space Science Reviews, 120, 143

\reference{} Beers, T., \& Christlieb, N.\ 2005, ARAA, 43, 531

\reference{} Belczynski, K., Kalogera, V., \& Bulik, T.\ 2002, \apj, 572, 407

\reference{} Belczynski, K., Perna, R., Bulik, T., Kalogera, V., Ivanova,
             N., \& Lamb, D.\ 2006, \apj, 648, 1110

\reference{} Belczynski, K., Bulik, T., Heger, A., \& Fryer, C.L.\ 2007, \apj,
             664, 986

\reference{} Belczynski, K., Kalogera, V., Rasio, F., Taam, R., Zezas, A.,
             Bulik, T., Maccarone, T., \& Ivanova, N. \ 2008a, \apjs, 174, 223

\reference{} Belczynski, K., Taam, R.E., Rantsiou, E., \& van der
             Sluys, M. 2008b, \apj, 682, 474

\reference{} Belczynski, K., O'Shaughnessy, R., Kalogera, V., Rasio,
             F., Taam, R. E., \& Bulik, T.\ 2008c, \apj, 680, L129

\reference{} Berger, E.\ 2007, \apj, 670, 1254

\reference{} Binney, J., \& Merrifield, M.\ 1998, Galactic Astronomy,
             Princeton University Press (chapters  6 \& 8)

\reference{} Bloom, J.S., Sigurdsson, S., \& Pols, O.\ 1999, \mnras, 305, 763

\reference{} Bloom, J., Frail, D., \& Kulkarni, S.\ 2003, \apj, 594, 674

\reference{} Bloom, J.S., Butler, N.R., \& Perley, D.A.\ 2008,
             astro-ph/0804.0965

\reference{} Butler, N.R., et al. 2007, \apj, 671, 656

\reference{} Cheng, K., \& Dai, Z.\ 1996,  Phys. Rev. Lett. 77, 1210

\reference{} Clark, J., Muno, M., Negueruela, I., Dougherty, P., Crowther,
             P., Goodwin, S., \& de Grijs, R.\ 2008, \aap, 477, 147

\reference{} Cusumano, G., et al.\ 2006, Nature, 440, 164

\reference{} Dai, Z., Wang, X., Wu, X., \&  Zhang, B. 2006, Science,
             311, 1127

\reference{} Dar, A., \& Rujula, A.\ 2004, Physics Reports, 405, 203

\reference{} Donaghy, T.Q. et al.\ 2006, astro-ph/0605570

\reference{} Duquennoy, A., \& Mayor, M.\ 1991, \aap, 248, 485

\reference{} Eichler, D., Livio, M., Piran, T., \& Schramm, D.\ 1989,
             Nature, 340, 126

\reference{} Erb, D., et al.\ 2006, \apj, 644, 813

\reference{} Fenimore, E., et al.\ 2004, Baltic Astronomy, 13, 301

\reference{} Fontana, A., et al.\ 2006, \aap, 459, 745

\reference{} Fox, D., et al.\ 2005, Nature, 437, 845

\reference{} Frail, D., et al.\ 2001, \apj, 562, L55

\reference{} Fryer, C.L., Woosley, S., \& Hartmann, D.H.\ 1999, \apj, 526, 152

\reference{} Fryer, C.L., \& Kalogera, V.\ 2001, \apj, 554, 548

\reference{} Fryer, C.L. et al. 2007, \pasp, 119, 1211

\reference{} Galama, T., et al.\ 1998, Nature, 395, 670

\reference{} Gehrels, N., et al.\ 2004, \apj, 611, 1005

\reference{} Gou, L., et al.\ 2004, \apj, 604, 508

\reference{} Greiner, J., et al.\ 2009, \apj, 693, 1610

\reference{} Grindlay, J.\ 2006, AIP Conf. Proc., 836, 631

\reference{} Grupe D., et al.\ 2006, \apj, 653, 462

\reference{} Guetta, G., Piran, T., \& Waxman, E.\ 2005, \apj, 619, 412

\reference{} Guetta, G., \& Piran, T.\ 2006, \aap, 453, 823

\reference{} Hartmann, D.H.\ 2008, New Astronomy Reviews, 52, 450

\reference{} Hartmann, D.H. et al.\ 2009, ``Tracing the Cosmic Star Formation
             History to its Beginnings: GRBs as Tools'', Astro2010 White Paper

\reference{} Heger, A., Fryer, C.L., Woosley, S.E., Langer, N., \& 
             Hartmann, D.H.\ 2003, \apj, 591, 288

\reference{} Heggie, D.\ C.\ 1975, \mnras, 173, 729

\reference{} Hjorth, J., et al.\ 2003, Nature, 423, 847

\reference{} Hogg, D.\ W.\ 2000, astro-ph/99055116

\reference{} Hopkins, A., \& Beacom, J.\ 2006, \apj,â 651, 142

\reference{} Hurley, J., Pols, O., \& Tout, C.\ 2000, \mnras, 315, 543

\reference{} Iye, M., et al.\ 2006, Nature, 443, 186 

\reference{} Janka, H.-T. \& Ruffert, M. 1996, \aap, 307, L33

\reference{} Kawai, N., et al.\ 2006, Nature, 440, 184

\reference{} King, A., Olsson, E., \& Davies, M.\ 2007, \mnras, 374, L34 

\reference{} Kistler, M., Yuksel, H., Beacom, J., \& Stanek, K.\ 2008, 
             \apj, 673, L119

\reference{} Kluzniak, W., \& Ruderman, M.\ 1998, \apj, 505, L113

\reference{} Kroupa, P., Tout, C.A., \& Gilmore, G.\ 1993, \mnras, 262, 545

\reference{} Kroupa, P., \& Weidner, C.\ 2003, \apj, 598, 1076

\reference{} Lamb, D. Q., \& Reichart, D.\ 2000, \apj, 536, 1

\reference{} Lamb, D. Q.,\ 2007, Philosophical Transactions of the Royal Society A: 
              365, 1363
\reference{} Lee, W.~H. \& Ramirez-Ruiz, E. 2007, New Journal of Physics, 9, 17

\reference{} Lloyd-Ronning, N.M., Fryer, C.L., Ramirez-Ruiz, E. 2002, ApJ, 574, 554

\reference{} MacFadyen, A.I., \& Woosley, S.E.\ 1999, \apj, 524, 262

\reference{} Mackey, J., Bromm, V., \& Hernquist, L.\ 2003, \apj, 586, 1 

\reference{} Metzger, B., Quataert, E., \& Thompson, T. 2008, \mnras, 
             385, 1455

\reference{} Metzger, B., Piro, A., \& Quataert, E.\ 2009, \mnras, 396, 304

\reference{} Nakar, E.\ 2007, Physics Reports, 442, 166

\reference{} Oechslin, R. \& Janka, H.-T. 2006, \mnras, 368, 1489

\reference{} O'Shea, B.W. \& Norman, M.L. 2007, \apj, 654, 660

\reference{} Paczynski, B.\ 1986, \apj, 308, L43

\reference{} Pei, Y., Fall, M., \& Hauser, M.\ 1999, \apj, 522, 604

\reference{} Perez-Ramirez, D., et al.\ 2008, \aap, submitted (arXiv:0810.2107)

\reference{} Piro, L., et al. 2009, Experimental Astronomy, 23, 67 

\reference{} Podsiadlowski, Ph., Mazzali, P.A., Nomoto, K., Lazzati, D., \&
             Cappellaro, E.\ 2004, \apj, 607, L17

\reference{} Popham, R., Woosley, S.E., \& Fryer, C.L.\ 1999, \apj, 518, 356

\reference{} Prochaska, J. X., Chen, H.-W., Dessauge-Zavadsky, M., and Bloom, 
             J. S. 2007, \apj, 666, 267

\reference{} Ripamonti, E., \& Abel, T.\ 2004, \mnras, 348, 1019

\reference{} Ruffini, R., et al.\ 2006, \apj, 645, 109

\reference{} Salvaterra, R., et al.\ 2009, Nature, 461, 1258

\reference{} Salvaterra, R., Guidorzi, C., Campana, S., Chincarini, G., \& 
             Tagliaferri, G.\ 2009, \mnras, 396, 299

\reference{} Sana, H., Gosset, E., Naze, Y., Rauw, G., \& Linder, N.\ 2008, 
             \mnras, 386, 447

\reference{} Savaglio, S., Glazebrook, K., \& LeBorgne, D 2009, \apj, 691,
             182 

\reference{} Schaerer, D.\ 2002, \aap, 382, 28

\reference{} Schady, P., et al.\ 2008, GCN Circ. 8217

\reference{} Schneider, R., Salvaterra, R, Ferrara, A., \& Ciardi, B.\ 
             2006, \mnras, 369, 825 

\reference{} Smith, B. et al.\ 2009, \apj, 691, 441

\reference{} Soderberg, A., et al.\ 2006, \apj, 650, 261

\reference{} Stamatikos, M., et al.\ 2008, GCN Circ. 8222

\reference{} Stanek, K., et al.\ 2003, \apj, 591, L17

\reference{} Strolger, L., et al.\ 2004, \apj, 613, 200

\reference{} Tanvir, N. R., et al.\ 2009, Nature, 461, 1254

\reference{} Tornatore, L., Ferrara, A., Schneider, R.\ 2007, 
             \mnras, 382, 945   

\reference{} Tremonti, C. et al.\ 2004, \apj, 613, 898

\reference{} Usov, V. 1992, Nature, 357, 472

\reference{} Virgili, F., Liang E., \& Zhang, B.\ 2009, \mnras, 
             392, 91 

\reference{} Vreeswijk, P., et al.\ 2008, GCN 8221

\reference{} Willot, C., et al.\ 2007, AJ, 134, 2435

\reference{} Woosley, S.\ 1993, \apj, 405, 273

\reference{} Woosley, S., \& Bloom, J.\ 2006, ARAA 44, 507

\reference{} Yoon, S.-C., Langer, N., \& Norman, C.\ 2006, 
             \aap, 460, 199

\reference{} Young, P. \& Fryer, C.L.\ 2007, \apj, 670, 584

\reference{} Zhang, B. Zhang, B.-B., Liang, Gehrels, N., 
             E.-W., Burrows, D.N., M\'esz\'aros, P.\ 2007, ApJ, 655, L25

\reference{} Zhang, B., et al.\ 2009, \apj, 703, 1696 


\end{references}
\end{document}